\pdfoutput=1

\documentclass{raa}
\usepackage{graphicx,times}
\usepackage{natbib}
\usepackage{amssymb,amsmath}
\bibpunct{(}{)}{;}{a}{}{,}

\begin{document}

\title{The nature of orbits in a prolate elliptical galaxy model
with a bulge and a dense nucleus}

\volnopage{ {\bf 2011} Vol.\ {\bf 11} No. {\bf 12}, 1449--1456}
   \setcounter{page}{1449}

\author{Nicolaos D. Caranicolas
\and Euaggelos E. Zotos}

\institute{Department of Physics,
Section of  Astrophysics, Astronomy and Mechanics,
Aristotle University of Thessaloniki
541 24, Thessaloniki, Greece;
{\it evzotos@astro.auth.gr}\\
\vs \no
   {\small Received 2011 June 3; accepted 2011 August 1}
}

\abstract{ We study the transition from regular to chaotic motion in
a prolate elliptical galaxy dynamical model with a bulge and
a dense nucleus. Our numerical investigation shows that stars with
angular momentum $L_z$ less {than }or equal to a critical value $L_{zc}$,
moving near the galactic plane{,} are scattered to higher $z$,
when reaching the central region of the galaxy, {thus }displaying chaotic
motion. An inverse square law relationship was found to exist
between the radius of the bulge and the critical value $L_{zc}$ of
the angular momentum. On the other hand, a linear relationship
exists between the mass of the nucleus and $L_{zc}$. The numerically
obtained results are explained using theoretical arguments.
Our study shows that there are connections between regular or
chaotic motion and the physical parameters of the system, such as
the star's angular momentum and mass, the scale length of the
nucleus {and} the radius of the bulge. The results are compared with
the outcomes of previous work. 
  }
\authorrunning{N. D. Caranicolas \& E. E. Zotos}
\titlerunning{The Nature of Orbits in a Prolate Elliptical Galaxy Model}
\maketitle

\section{Introduction}

Today astronomers believe that the intrinsic shapes of most
elliptical galaxies are either oblate or prolate. It is also true
that spherical galaxies are very rare {but} triaxial elliptical
galaxies do exist (see \citealt{Statler1994, Ryden1996, Alam2002,
Vincent2005}). On the other hand, there are observational data
indicating the presence of a black hole or a dense massive nucleus
in the central parts of elliptical galaxies (\citealt{Statler2004}).

All of the above gives us the opportunity to construct a dynamical
model for an elliptical galaxy hosting a dense nucleus, in order
to use it for the study of the global properties of motion in
these stellar systems. In order to describe the motion in a
prolate elliptical galaxy we use the potential
\begin{equation}
V_{\rm t} = \frac{\upsilon_0^2}{2}\ln\left(r^2 + \alpha z^2 +c_{\rm
b}^2 \right) - \frac{M_{\rm n}}{\sqrt{r^2 + z^2 +c_{\rm n}^2}}
 = V_{\rm g} + V_{\rm n}. \ \ \
\end{equation}
Our model consists of two components. The first component describes
a prolate elliptical galaxy while the second is the potential of a
dense spherical nucleus. Here $(r, z)$ are the usual cylindrical
coordinates, $\upsilon_0$ is used for 
 consistency of 
galactic units, $0.2 \leq \alpha <1$ is the flattening parameter,
{and} $c_{\rm b}$ is the radius of the bulge component. Furthermore,
$M_{\rm n}$ is the mass and $c_{\rm n}$ is the scale length of the
nucleus.

In an earlier paper (\citealt{Caranicolas1991}, hereafter Paper I)
we have studied the transition from regular to chaotic motion in a
disk galaxy model with a dense nucleus. There we found that stars,
moving in the $(r,z)$ plane with values of angular momentum $L_z$
less than or equal to a critical value $L_{zc}$, are scattered to
higher $z$ upon encountering the dense nucleus, thus displaying
chaotic motion. We also found relationships connecting the
physical parameters of the system with chaos. In the present
paper, we shall focus our study 
 on the transition from regular to
chaotic motion in 
  model (1), which describes a prolate
elliptical galaxy hosting a dense nucleus. Our aim is: (i) to look
for relationships between the physical parameters and chaos, (ii)
to explain the numerically found relationships using a
semi-theoretical approach and (iii) to compare the present results
with the outcomes found in Paper I.

In this research we use the well known system of galactic units,
where the unit of length is $1$\,kpc, the unit of mass is $2.325
\times 10^7\,M_{\odot}$ and the unit of time is $0.97748 \times
10^8$\,yr. The velocity and the angular velocity units are
$10$~km~s$^{-1}$ and $100$~km~s$^{-1}$~kpc$^{-1}$ respectively,
while $G$ is equal to unity. The energy unit (per unit mass) is
$100$~km$^2$~s$^{-2}$. In the above units we use the values:
$\upsilon_0=10$ while $0 \leq M_{\rm n} \leq 150$, $0.75 \leq c_{\rm
b} \leq 1.25$ and $0.1 \leq c_{\rm n} \leq 0.25$.
{Since} the total potential $V_{\rm t} = V_{\rm
t}\left(r, z\right)$ is axially symmetric and the $L_z$ component of
the angular momentum is conserved{,} we use the effective potential
\begin{equation}
V_{\rm eff}=\frac{L_z^2}{2r^2}+V_{\rm t}\left(r,z\right),\ \ \
\end{equation}
in order to study the motion in the meridian $(r-z)$ plane. The equations of motion are
\begin{eqnarray}
\dot{r} = p_r, \ \ \ \dot{z} = p_z, \nonumber\\
\dot{p_r} = -\frac{\partial \ V_{\rm eff}}{\partial r}, \ \ \
\dot{p_z} = -\frac{\partial \ V_{\rm eff}}{\partial z}, \ \ \
\end{eqnarray}
and the corresponding Hamiltonian is written{ as}
\begin{equation}
H=\frac{1}{2}\left(p_r^2 + p_z^2 \right) + V_{\rm
eff}\left(r,z\right) = E, \ \ \
\end{equation}
where $p_r${ and} $p_z$ are the momenta per unit mass
conjugate to $r$ and $z${ respectively}, while $E$ is the numerical value of the
Hamiltonian. Equation~(4) is an integral of motion, which indicates
that the total energy of the particle is conserved. Orbit
calculations are based on the numerical integration of the equations
of motion (3), which {were} made using a Bulirsh-St\"{o}er routine, {to}
double precision. The accuracy of the calculations was checked by
the constancy of the energy integral, which was conserved up to the
twelfth significant figure.

The paper is organized as follows. In Section~2 we present
numerical results for the potential of a non-active galaxy, that
is when $M_{\rm n}=0$. Furthermore, the numerical relationship is
explained using some semi-theoretical arguments. In Section~3, we
investigate numerically the case when the dense nucleus is
present. Some semi-theoretical argument{s} are also presented, in
order to explain the numerical outcomes. In Section~4 a comparison
with earlier work is {given} and discussion and conclusions of
this research are presented.

\section{Results when the massive dense nucleus is not present}

\begin{figure}[]
\centering

\resizebox{0.6\hsize}{!}{\rotatebox{0}{\includegraphics*{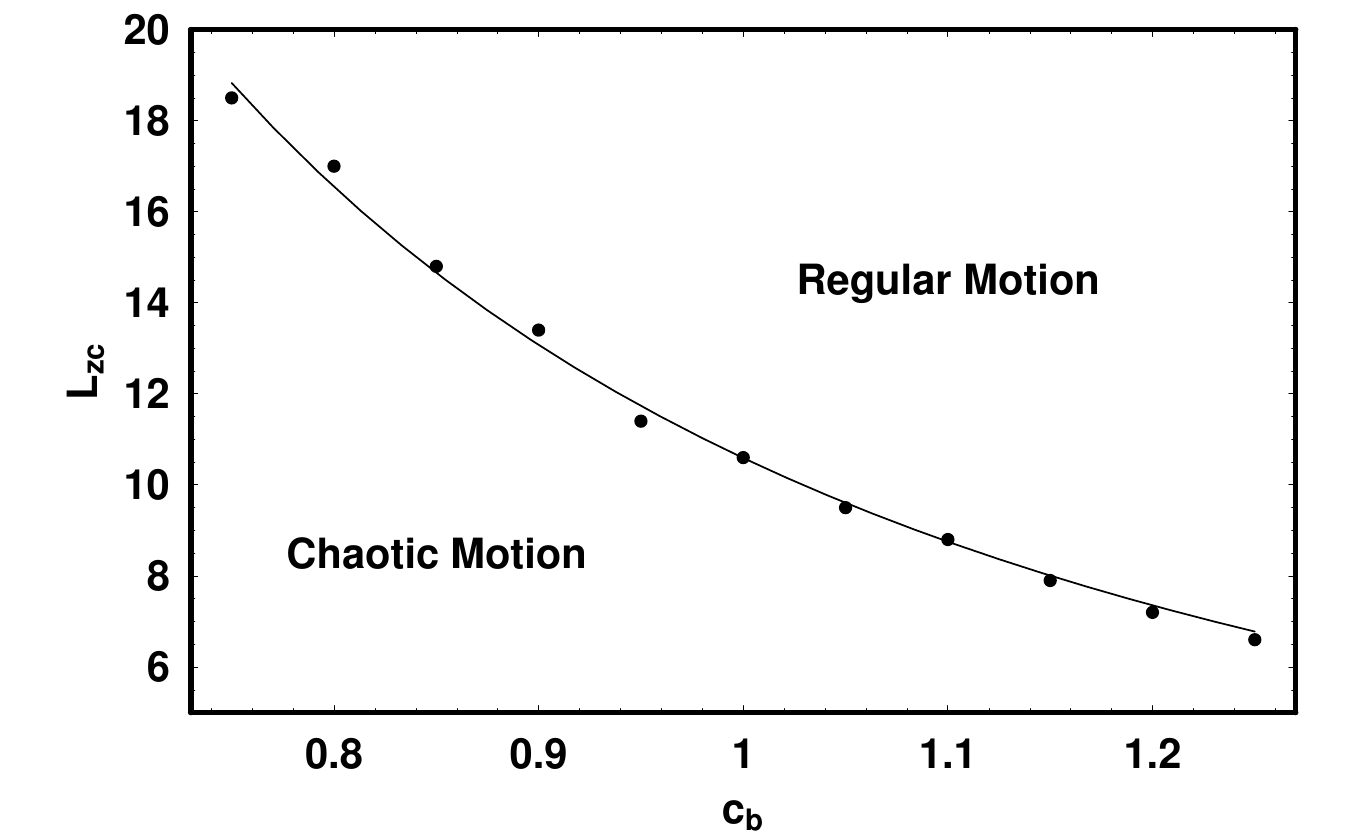}}}
\begin{minipage}{110mm}

\caption{
Plot of $L_{zc}$ vs. $c_{\rm b}$. The values of the parameters are
given in the text.\label{fig1}}\end{minipage}
\end{figure}

In this Section we shall study the behavior of orbits when the
dense massive nucleus is not present, that is when $M_{\rm n}=0$.
Figure~\ref{fig1} 
shows the numerical
relationship between the critical value $L_{zc}$ of the angular
momentum and the radius of the bulge $c_{\rm b}$. Orbits were
started near $r_0=r_{\rm max}$, with $z_0 = p_{r0}=0$, while the
value of $p_{z0}$ is always found from the energy integral (4).
The value of $r_{\rm max}$ is the maximal root of equation
\begin{equation}
\frac{L_z^2}{2r^2}+\frac{1}{2}\ln\left(r^2 + c_{\rm b}^2 \right)=E,
\ \ \
\end{equation}
which was found numerically. Dots represent the numerical
values while the solid line is the best fit which is an inverse
square law, represented by the equation
\begin{equation}
L_{zc}=\frac{10.5906}{c_{\rm b}^2}. \ \ \
\end{equation}
The value of $\alpha$ is 0.2, while the value of $E$ is 231. Orbits
with values of the parameters on and below the line are chaotic,
while orbits on the upper side of the line are regular.

\begin{figure*}[]
\centering

\includegraphics[width=70mm]{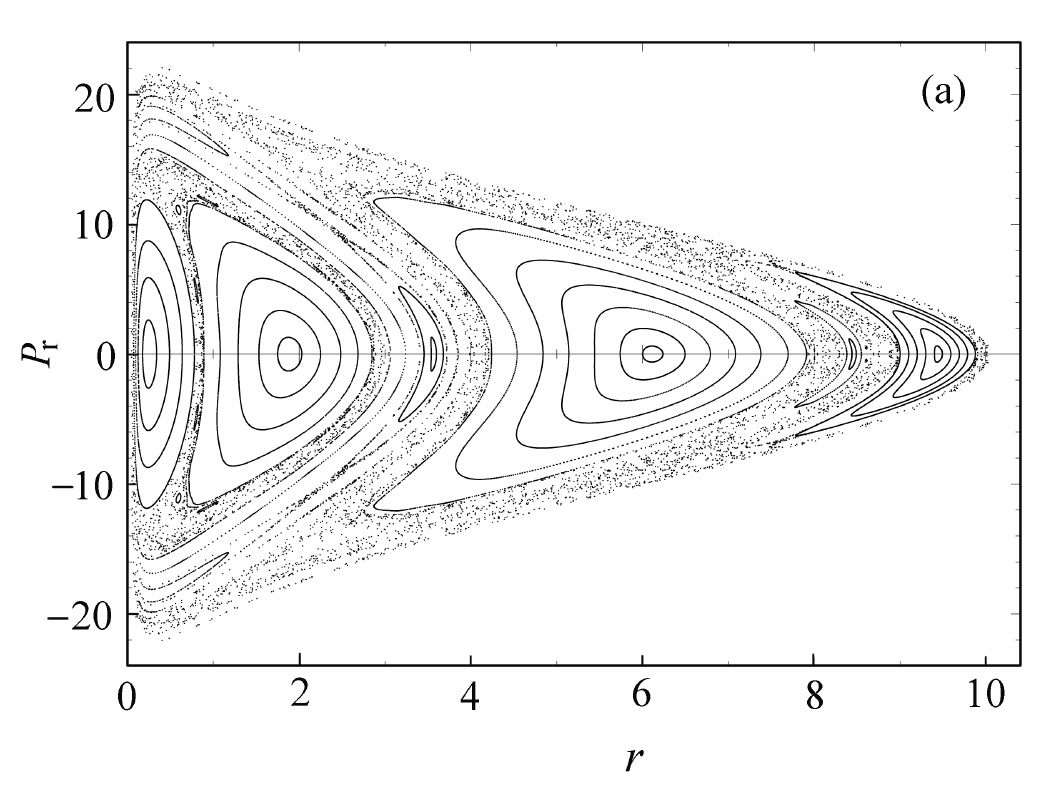}~~~
\includegraphics[width=70mm]{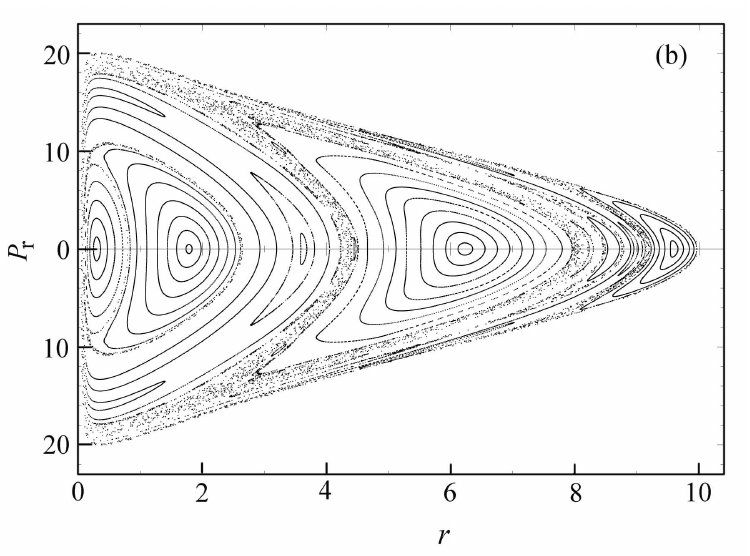}


\vspace{-4mm}
\caption{\baselineskip 3.6mm 
  $(r,p_r)$ phase plane
when the dense nucleus is absent. (a) 
  $c_{\rm b}=0.75$ and (b) 
  $c_{\rm b}=1.25$. The values of
all other parameters are given in the text. \label{fig2}}
\end{figure*}

\begin{figure*}[]


\includegraphics[width=70mm]{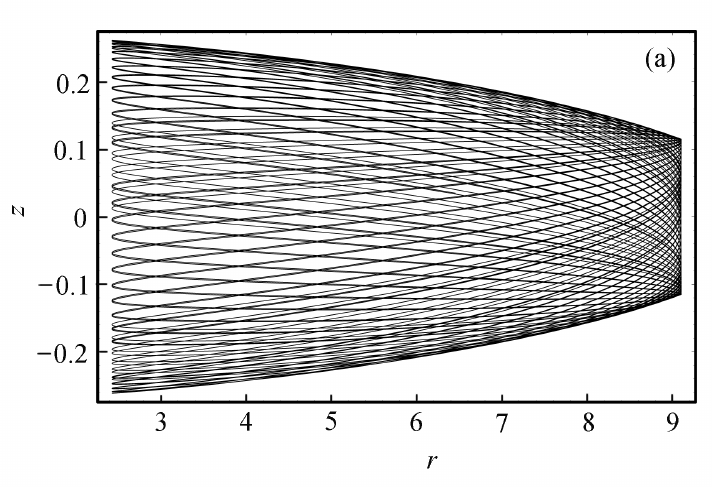}
\includegraphics[width=70mm]{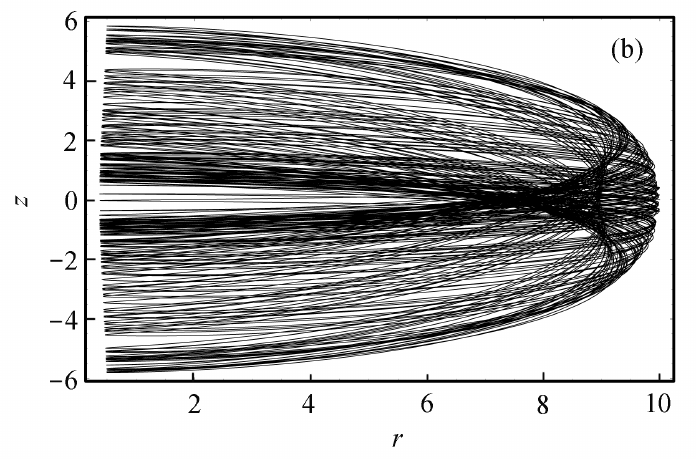}

\vspace{-4mm}

\caption{\baselineskip 3.6mm  
Orbits when the dense nucleus is not present. (a) 
A regular orbit which stays near the galactic plane. 
(b) 
 A chaotic orbit which is scattered into high values of
$z$. See the text for details.\label{fig3}}
\end{figure*}

Figure~\ref{fig2} 
shows the Poincar\'{e} $(r,p_r), z=0, p_z>0$ phase plane when
$\alpha=0.2, L_z=1, E=213$, while the value of $c_b$ is 0.75 in
Figure~\ref{fig2}(a) and 1.25 in Figure~\ref{fig2}(b). As one can
see, the pattern is similar in both figures with regions of
regular and chaotic motion. A more detailed inspection shows that
the extent of the chaotic zone is larger in the case when
$c_b=0.75$, which is when we have a galaxy with a denser bulge.
Figure~\ref{fig3} 
  shows two orbits when $\alpha=0.2, c_b=0.9,
E=231$. The value of the angular momentum in the orbit shown in
Figure~\ref{fig3}(a) is $L_z=40$ and the initial conditions are:
$r_0=9.1, z_0 = p_{r0}=0$, while the value of $p_{z0}$ is always
found from the energy integral (4). Note that the orbit is regular
and stays very close to the galactic plane. On the contrary, the
orbit shown in Figure~\ref{fig3}(b) has a value of $L_z=8$ and
initial conditions: $r_0=10.0, z_0 = p_{r0}=0$. The orbit is
chaotic and it is scattered off the galactic plane, displaying
high values of $z$. Both orbits were calculated for a time period
of 200 time units.

Let us now {begin} to use semi-analytical arguments in order to
explain the numerically found relationship of Figure~\ref{fig1}.
The lines of arguments are similar to those used in Paper I. As
the test particle {approaches} very close to the center of a
galaxy, there is a change in its momentum in the $z$ direction
given by the equation
\begin{equation}
m\Delta p_z = \langle F_z \rangle \Delta t. \ \ \
\end{equation}
Here $m$ is the mass of the test particle{,} $\langle F_z \rangle$
is the total average force acting in the $z$ direction and $\Delta
t$ is the duration of the encounter. Empirical evidence shows that
the test particle's rise proceeds cumulatively in each case, 
and increases a little more with each successive pass from the
center rather than with a single ``violent" encounter. It is
observed that the test particle gains considerable height after $n
(n>1)$ crossings, when the total change in the momentum in the $z$
direction is {on the} order of $m \upsilon_\phi$, where
$\upsilon_\phi$ is the tangential velocity of the test particle
near the center at a distance $r=\langle r_0 \rangle \simeq
\langle z_0 \rangle \ll 1$. Therefore we write
\begin{equation}
m\sum\limits_{i=1}^{n}{\Delta {{p}_{zi}}}\approx \langle
{{F}_{z}}\rangle \sum\limits_{i=1}^{n}{\Delta {{t}_{i}}}. \ \ \
\end{equation}
Setting
\begin{eqnarray}
m\sum\limits_{i=1}^{n}{\Delta {{p}_{zi}}}
 &=& m{{\upsilon }_{\phi }}=\frac{m{{L}_{zc}}}{\langle {{r}_{0}}\rangle}\,, \nonumber \\
\sum\limits_{i=1}^{n}{\Delta {{t}_{i}}} &=& {{T}_{c}}, \ \ \
\end{eqnarray}
in Equation~(8) we find
\begin{equation}
\frac{m{{L}_{zc}}}{\langle{{r}_{0}}\rangle}\,\approx \,\langle
{{F}_{z}}\rangle {{T}_{c}}. \ \ \
\end{equation}

The force acting in the $z$ direction for a test particle of unit mass $(m=1)$ is
\begin{equation}
{{F}_{z}}=\frac{-\upsilon _{0}^{2} \alpha z}{{{r}^{2}}+ \alpha
{{z}^{2}}+c_{\rm b}^{2}}. \ \ \
\end{equation}
Remember that $r=\langle r_0 \rangle \simeq \langle z_0 \rangle
\ll 1$, therefore $\langle r_0^2 \rangle \simeq \langle z_0^2
\rangle \ll c_{\rm b}^2$. Keeping only the linear terms in
$\langle r \rangle$ and $\langle z \rangle$ in Equation~(11) and
taking the absolute value of the $F_z$ force, we find from
relationship (10) that
\begin{equation}
{{L}_{zc}}\,\approx \,\frac{\upsilon _{0}^{2} \alpha {{\left(
\langle {{r}_{0}}\rangle \right)}^{2}}}{c_{\rm b}^{2}}{{T}_{c}}. \ \
\
\end{equation}
{Since} $T_c$ was observed to be the same when $0.75
\leq c_{\rm b} \leq 1.25$, we can set $k_1=\upsilon _0^2 \alpha
\left(\langle r_0 \rangle \right)^2 T_c$ and obtain
\begin{equation}
L_{zc} \approx \frac{k_1}{c_{\rm b}^2}, \ \ \
\end{equation}
which explains the numerical relationship of
Figure~\ref{fig1}. The authors would like to make {it
}clear that relation~(13) does not reproduce the
numerical results shown in Figure~\ref{fig1}, but {only }shows
the form of the relationship.

\section{Results when the massive nucleus is present}

\begin{figure}
\centering
\resizebox{0.6\hsize}{!}{\rotatebox{0}{\includegraphics*{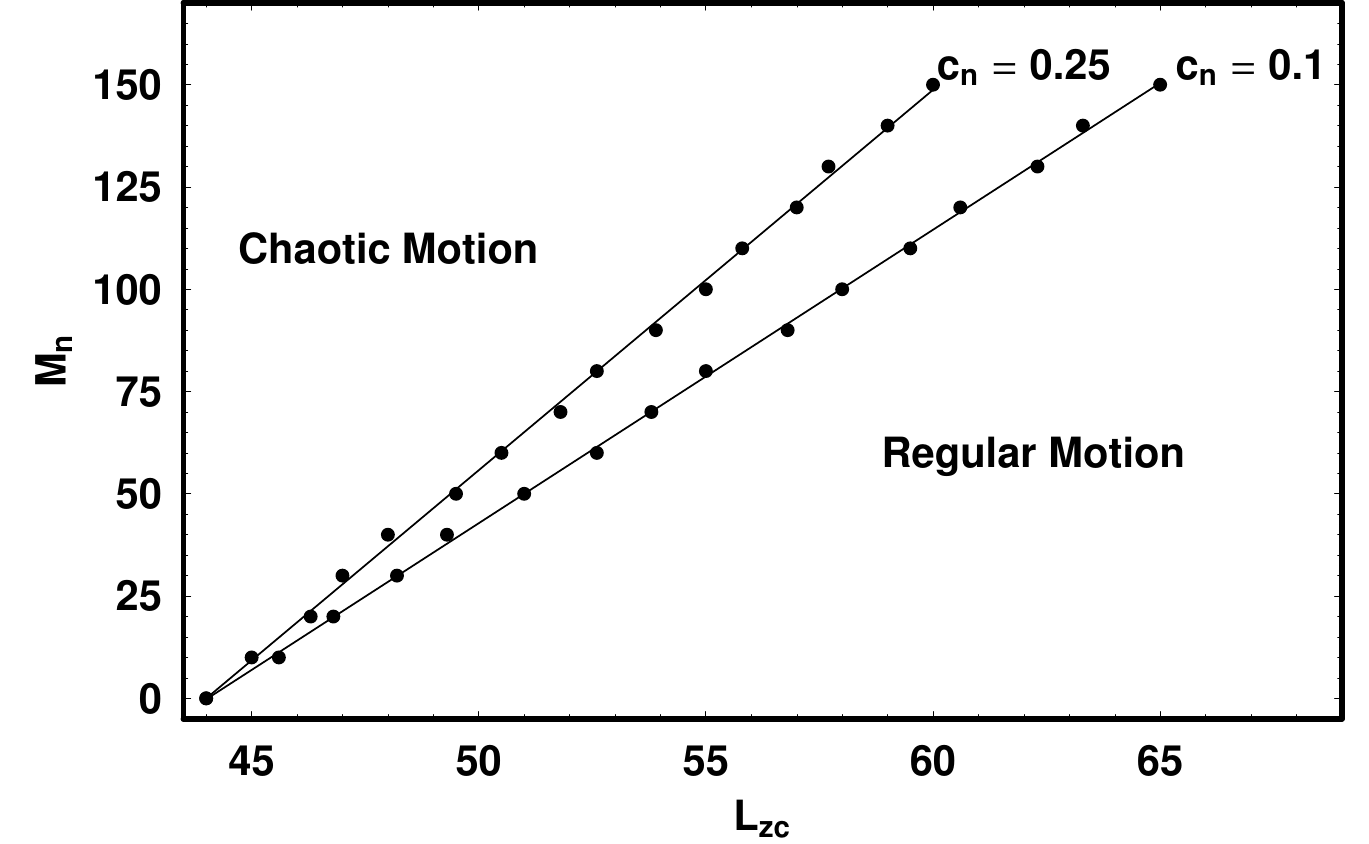}}}

\begin{minipage}{110mm}
\caption{
Plot of $M_{\rm n}$ vs. $L_{zc}$. The values of the parameters are
given in {the }text.\label{fig4}}\end{minipage}

\vs

\includegraphics[width=70mm]{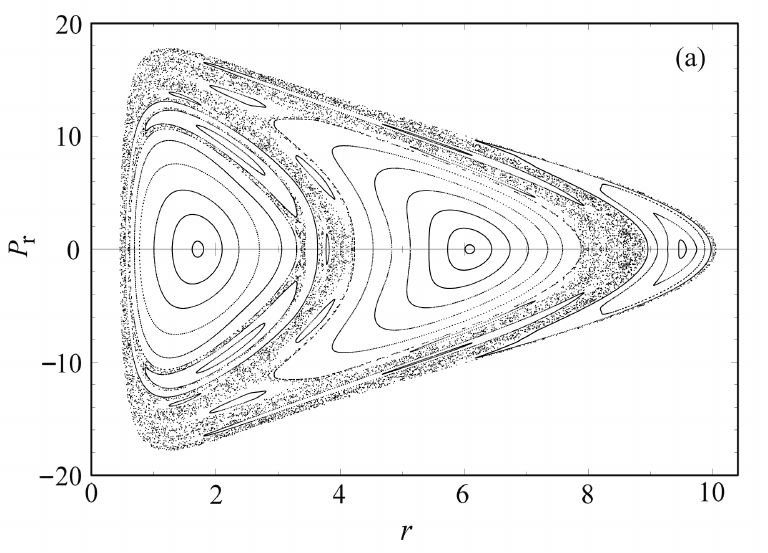}
\includegraphics[width=70mm]{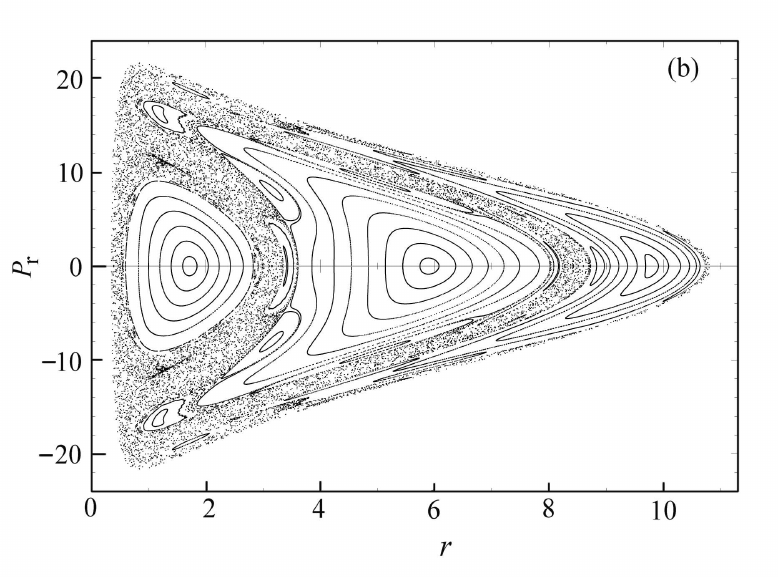}

\vspace{-4mm}
\caption{\baselineskip 3.6mm 
 $(r,p_r)$ phase plane when
the dense nucleus is present. (a) 
$M_{\rm n}=20$ and (b) 
$M_{\rm n}=100$. The values of all other parameters are given in
the text.\label{fig5}}
\end{figure}

We investigate the case when we have an active galaxy, that is
when the dense massive nucleus is present. Figure~\ref{fig4}
   shows a numerical relationship between the
critical value $L_{zc}$ of the angular momentum and the mass of
the nucleus $M_{\rm n}$ for two values of $c_{\rm n}$. The
procedure to obtain the results shown in Figure~\ref{fig4} is
similar to that followed in Figure~\ref{fig1}. The value of
$\alpha$ is 0.5, $c_{\rm b}=1.2$, while the value of $E$ is 227.
Here we see a straight line. Orbits with values of the parameters
on the left side of the line{,} including the line{,} are chaotic,
while orbits on the right side of the line are regular. It is
interesting to note that the extent of the chaotic region is
larger when the value of $c_{\rm n}$ is smaller, {which} is when
we have a denser nucleus.

\begin{figure}[]

\centering

\includegraphics[width=72mm]{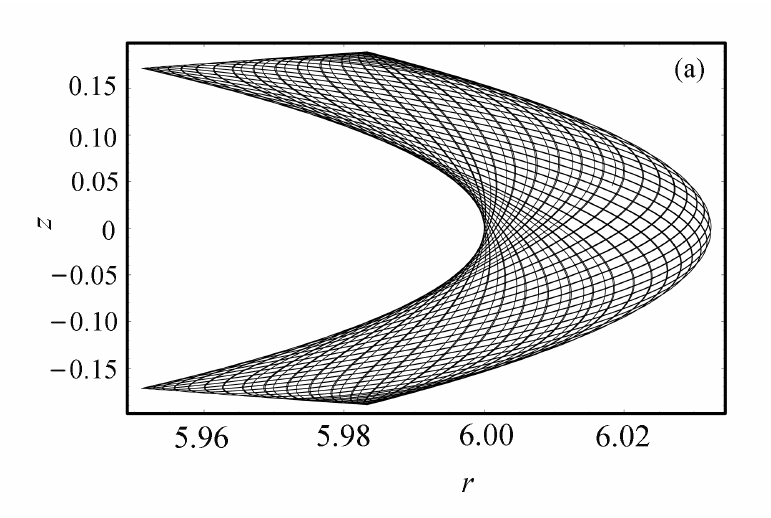}
\includegraphics[width=68mm]{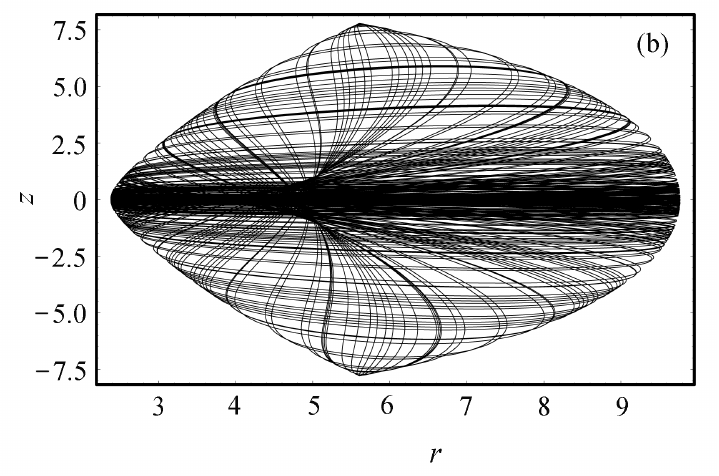}

\vspace{-4mm}
\caption{\baselineskip 3.6mm 
 Orbits when the dense nucleus
is present. (a) A regular orbit which stays near the galactic
plane. 
  (b) A chaotic orbit which is scattered into high values
of $z$. See {the }text for details.\label{fig6}}

\vs\vs \centering
\resizebox{0.5\hsize}{!}{\rotatebox{0}{\includegraphics*{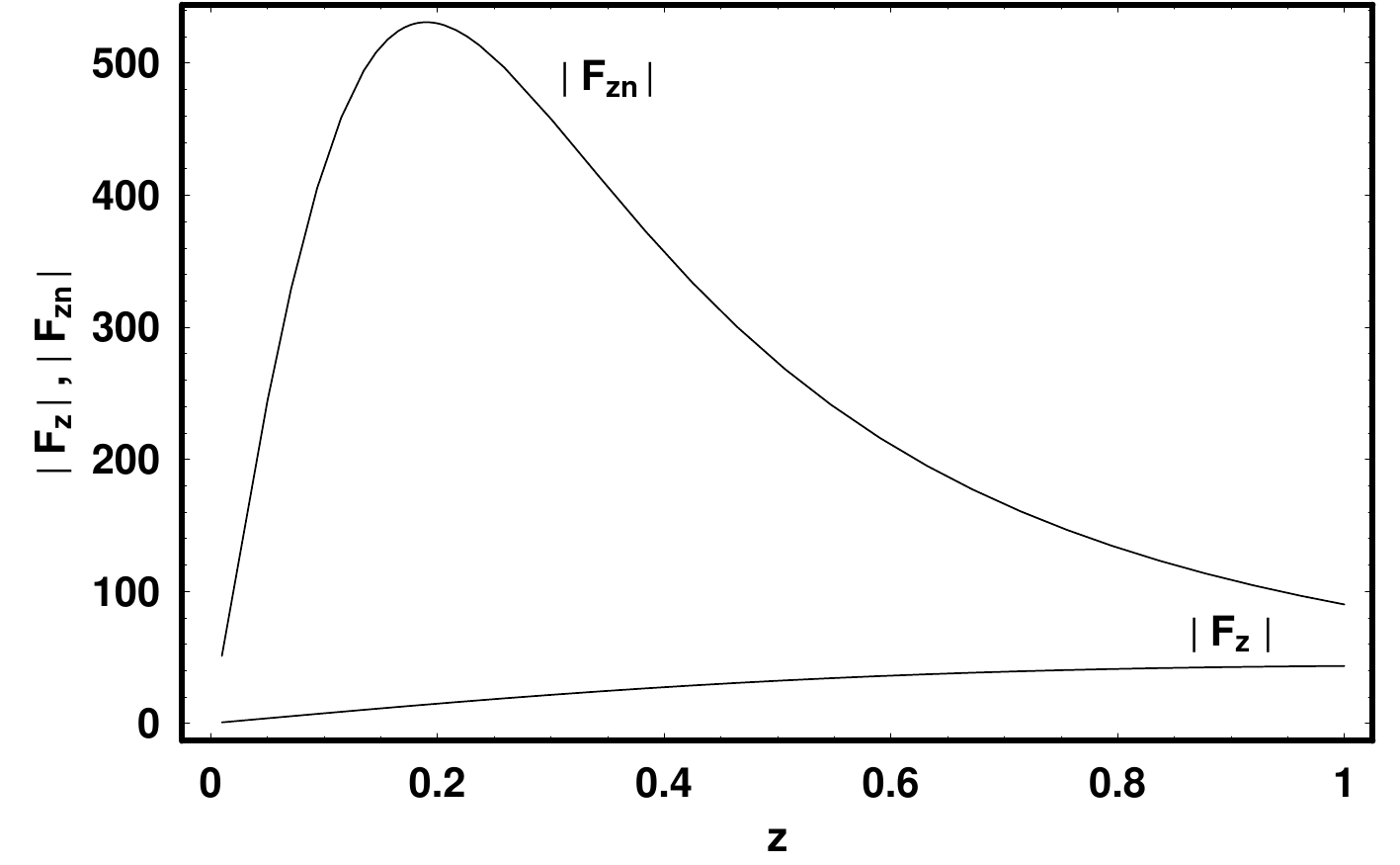}}}

\vspace{-4mm}
\caption{
Plot of $|F_z|$ and $|F_{zn}|$ vs. $z$. The values of all the
other parameters are given in {the }text.\label{fig7}}
\end{figure}

Figure~\ref{fig5}(a)--(b) 
 shows the
Poincar\'{e} $(r,p_r), z=0, p_z>0$ phase plane when: $\alpha=0.2,
L_z=10, E=230, c_{\rm b} =1.2${;} the value of $M_{\rm n}$ is 20
in Figure~\ref{fig5}(a) and 100 in Figure~\ref{fig5}(b). In both
cases, we observe regular regions together with large chaotic
regions. Some small islands are also present indicating secondary
resonances. It is evident that the area covered by chaotic orbits
is larger in the case of $M_{\rm n}=100$, {which} is when we have
a more massive nucleus. In Figure~\ref{fig6}(a)--(b)
  we can see two orbits when $\alpha=0.5, c_{\rm
b}=1.2,  E=227$. The value of the angular momentum in the orbit
shown in Figure~\ref{fig6}(a) is $L_z=60$, while the value of
$M_{\rm n}$ is 25 and $c_{\rm n}=0.25$. The initial conditions are
$r_0=6.0, z_0 = p_{r0}=0$, while the value of $p_{z0}$ is always
found from the energy integral (4). Note that the orbit is
regular, and stays very close to the galactic plane. On the other
hand, the orbit shown in Figure~\ref{fig6}(b) has a value of
$L_z=45$, while the value of $M_{\rm n}$ is 120 and $c_{\rm
n}=0.25$. Initial conditions are: $r_0=9.768, z_0 = p_{r0}=0$. The
orbit is chaotic and it is scattered off the galactic plane{,}
displaying high values of $z$. Both orbits were calculated for a
time period of 200 time units.

The linear relationship of Figure~\ref{fig4} can be obtained using
semi-analytical arguments. As the test particle approaches the
nucleus it experiences a strong vertical force, due to the
presen{ce} of the dense nucleus.

Figure~\ref{fig7} 
shows a plot of{ the} $F_z$ force
as well as the nuclear $F_{zn}$ force as a function of $z$, near
the nucleus when $r_0=0.1$. The values of the parameters are
$\alpha=0.5, \upsilon_0=10, c_{\rm b}=0.8, M_{\rm n}=100, c_{\rm
n}=0.25$. We see that the nuclear force is about 25 times as
strong as $F_z$. Because of this strong force, there is a change
in its momentum in the $z$ direction given by the equation
\begin{equation}
m\Delta p_z = \langle F_{zn} \rangle \Delta t, \ \ \
\end{equation}
where $m$ is the mass of the test particle{,} $\langle F_{zn} \rangle$
is the average nuclear force acting in the $z$ direction, while
$\Delta t$ is the duration of the encounter. Here again, our
numerical results show that the test particle goes to higher $z$
after $n (n>1)$ crossings, when the total change in the momentum in
the $z$ direction is {on the} order of $m \upsilon_\phi$, where
$\upsilon_\phi$ is the tangential velocity of the test particle near
the center, at a distance $r_0 \simeq z_0 \simeq c_{\rm n}$.
Therefore we write
\begin{equation}
m\sum\limits_{i=1}^{n}{\Delta {{p}_{zi}}} \approx \langle
{{F}_{zn}}\rangle \sum\limits_{i=1}^{n}{\Delta {{t}_{i}}}. \ \ \
\end{equation}
If we set
\begin{eqnarray}
m\sum\limits_{i=1}^{n}{\Delta {{p}_{zi}}} &=& m{{\upsilon }_{\phi }}=\frac{m{{L}_{zc}}}{\langle {{r}_{0}}\rangle}, \nonumber \\
\sum\limits_{i=1}^{n}{\Delta {{t}_{i}}} &=& {{T}_{e}}, \nonumber \\
m &=& 1, \nonumber \\
\langle F_{zn} \rangle &=& \frac{c_{\rm n} M_{\rm n}}{\left(3c_{\rm
n}^2 \right)^{3/2}}, \ \ \
\end{eqnarray}
in Equation (15) we obtain
\begin{equation}
M_{\rm n} \approx k_2 L_{zc} c_{\rm n}, \ \ \
\end{equation}
where $k_2=\sqrt{27}/T_e$. Note that the values of $T_e$ were
observed to be about the same when $0 < M_n \leq 150$. Equation
(17) explains the numerically found relationship given in
Figure~\ref{fig4}.

\section{Discussion}

Today it is well{-}known that there are luminous elliptical
galaxies hosting dense massive nuclei (see \citealt{Urry2000}).
During the {past few} years{,} a large amount of observational
data {has given} a better and more detailed picture of these
active galaxies (\citealt{Barth2002, Costamante2002, Falomo2002,
Vagnetti2003, Falcone2004, Heidt2004, Bramel2005, Nieppola2006,
Zheng2007}).

In this article we have studied the transition from regular to
chaotic motion in a prolate elliptical galaxy model. Two cases
were studied, which are the case 
 where the nucleus was absent and
the case where we have an active galaxy hosting a dense massive
nucleus. In the first case we found that an inverse square law
relationship exists between the radius of the bulge of the galaxy
and the critical angular momentum when all other parameters are
kept fixed. This relationship was also reproduced, using some
semi-analytical arguments.

On the other hand, it was observed that for larger values of the
radius of the bulge{,} for $c_{\rm b} \geq 1.5$, the chaotic
regions when the dense nucleus is absent, if any, are negligible
when $0.2 \leq \alpha < 1$. This result is in agreement with the
result found in Paper I, where no chaos was observed when $c_{\rm b}
> 1.3$ and this result was independent of the mass of the bulge.
Thus, our results suggest that chaos is observed in disk and prolate
elliptical galaxies when a dense bulge is present.

Interesting results are obtained when 
 a dense nucleus is present.
In this case the numerically found relationship connecting $L_{zc}$
and $M_{\rm n}$ is linear when all other parameters are kept
constant. Furthermore, the linear relationship depends on the value
of $c_{\rm n}$, in such a way that the extent of the chaotic region
is larger when $c_{\rm n}$ is smaller{,} {which} is when the nucleus has a
higher density. The linear relationship obtained by the numerical
integration of the equations of motion was also found using some
semi-theoretical arguments, together with numerical evidence.

The present investigation shows that the results obtained are very
similar to those obtained for disk galaxies in Paper I. Therefore,
we can say that low angular momentum stars approaching a dense
massive nucleus are deflected to higher $z${, thus} displaying chaotic
motion. The similarity of the results obtained for different
galactic models suggests that it is the strong vertical force near
the dense nucleus that is responsible for this scattering combined
with the star's low angular momentum, which allows the star to
approach the dense nucleus.

Some of the latest discoveries obtained from observational
astronomy show that supermassive black holes, up to a billion
solar masses, inhabit the centers of all massive spheroidal
galaxies, independent of their visible activity; the supermassive
black holes are often quiescent with regard to their own radiation
but always show dynamical behavior. We believe that, with new data
from active galaxies, astronomers will be able{ to construct
better dynamical models} in the near future, in order to study the
properties of motion in galaxies and to find interesting
relationships connecting chaos with the physical parameters of
these stellar systems.

\begin{acknowledgements}
Useful suggestions and comments from an anonymous referee are gratefully acknowledged.
\end{acknowledgements}

\end{document}